\documentclass[aip,rsi,preprint,groupedaddress,showmsc,superscriptaddress,floatfix,pra,aps,showpacs,twocolumn]{revtex4}
\usepackage{amsmath,amsfonts,amssymb,caption,hyperref,color,epsfig,graphics,graphicx,latexsym,mathrsfs,revsymb,theorem,url,verbatim,epstopdf,cleveref}
\hypersetup{colorlinks,linkcolor={blue},citecolor={blue},urlcolor={red}}

\newenvironment{proof}{\noindent \textbf{{Proof.~} }}

\newtheorem{definition}{Definition}
\newtheorem{proposition}[definition]{Proposition}
\newtheorem{lemma}[definition]{Lemma}

\newtheorem{theorem}[definition]{Theorem}
\newtheorem{corollary}[definition]{Corollary}
\newtheorem{conjecture}[definition]{Conjecture}

\newtheorem{remark}[definition]{Remark}
\newtheorem{example}[definition]{Example}
\newtheorem{question}[definition]{Question}

\def\bcj{\begin{conjecture}}
\def\ecj{\end{conjecture}}
\def\bcr{\begin{corollary}}
\def\ecr{\end{corollary}}
\def\bd{\begin{definition}}
\def\ed{\end{definition}}
\def\bea{\begin{eqnarray}}
\def\eea{\end{eqnarray}}
\def\bem{\begin{enumerate}}
\def\eem{\end{enumerate}}
\def\bex{\begin{example}}
\def\eex{\end{example}}
\def\bim{\begin{itemize}}
\def\eim{\end{itemize}}
\def\bl{\begin{lemma}}
\def\el{\end{lemma}}
\def\bpf{\begin{proof}}
\def\epf{\end{proof}}
\def\bpp{\begin{proposition}}
\def\epp{\end{proposition}}
\def\bqu{\begin{question}}
\def\equ{\end{question}}
\def\br{\begin{remark}}
\def\er{\end{remark}}
\def\bt{\begin{theorem}}
\def\et{\end{theorem}}
\def\btb{\begin{tabular}}
\def\etb{\end{tabular}}

\def\tr{\mathop{\rm Tr}}
\begin{document}

\title{Faithful coherent states}

\author{Jun Li}
\thanks{Corresponding author: junlimath@buaa.edu.cn}
\affiliation{LMIB(Beihang University), Ministry of Education, and School of Mathematical Sciences, Beihang University, Beijing 100191, China}
\author{Lin Chen}
\thanks{Corresponding author: linchen@buaa.edu.cn}
\affiliation{LMIB(Beihang University), Ministry of Education, and School of Mathematical Sciences, Beihang University, Beijing 100191, China}
\affiliation{International Research Institute for Multidisciplinary Science, Beihang University, Beijing 100191, China}

\begin{abstract}
We propose the notion of faithful coherent states based on the fidelity-based coherence witness. The criterion for detecting faithful coherent states can be restricted to a subclass of fidelity-based criterion under unitary transformations for single and bipartite systems. We can realize these unitary transformations by using quantum gates and circuits, and establish the connection of faithful coherence states and coherence distillation, maximum relative entropy of coherence.
\end{abstract}
\maketitle

\section{INTRODUCTION}
\label{intro}
\begin{figure}[h]
\centering
\includegraphics[width=3.5in]{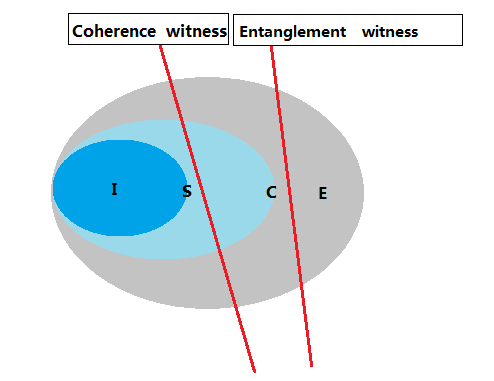}
\caption{All entanglement witnesses are also coherence witnesses for a fixed basis in multipartite quantum systems, where $\textbf{I}$ is the set of all incoherent states, $\textbf{S}$ is the set of all separable states, $\textbf{C}$ is the set of all coherent states and $\textbf{E}$ is the set of all entangled states. The relation among these sets is $\textbf{I} \subseteq \textbf{S}$, $\textbf{E} \subseteq \textbf{C}$, $\textbf{S} = \textbf{I}\cup (\textbf{C}-\textbf{E})$ and $\textbf{C}=(\textbf{S}-\textbf{I})\cup \textbf{E}$.}
\label{fig:cwew}
\end{figure}

Quantum coherence \cite{TBMC,ASGA} plays a central role in quantum physics and information science. The detection and quantifying of coherence have been widely investigated \cite{TBMC,ASGA,XDYD,MLHX}, which is as important as those of entanglement as one of quantum resource theory. Entanglement witnesses are fundamental tools in quantum entanglement theory, which are observables that completely separate separable states from quantum states and allow us to detect entanglement physically \cite{MHPH,BMTL,MBME,GTOG,SXYN,LCYX}. Inspired by the entanglement witness, coherence witness was first introduced in \cite{CNTR}, which is one of experimentally implementable ways to detect coherence by measuring the expectation values. Notably, the entanglement witnesses are also kinds of coherence witnesses for a fixed basis in multipartite quantum systems. The reason is that the entangled states can not be diagonal form, they can only be coherent states. We can describe the relation between entanglement and coherence in FIG \ref{fig:cwew}. In \cite{MWBD}, the authors proposed the faithful states whose entanglement can be characterized by using fidelity and provided the criterion for distinguishing unfaithful and faithful states. Furthermore, the authors in \cite{OGYM} provided a structural result on such entanglement faithful states for bipartite systems, which established connections to computational complexity and simplify several results in entanglement theory. Recently, the phenomenon of entangled faithfulness was also studied experimentally. \cite{GRDE,XMHW}. Coherence is another quantum resource theory. Inspired by entanglement faithful, we can consider the properties of coherence faithful. As far as we know, the notion and properties of faithful coherent states have not been studied yet.

In this paper, we propose faithful coherence based on coherence witness. The coherent witness here is the fidelity-based witness, which can be constructed and generalized to multipartite systems \cite{BHWZ,ZMZZ}. First, we prove that the criterion for detecting faithful coherence states can be restricted to a special subclass of fidelity-based criterion for single and bipartite systems in Theorems \ref{Theorem1} and \ref{Theorem2}. Then, we study the experimental implementation of unitary transformation in the theorems. We can realize these unitary transformations by using quantum gates and circuits, which are stable and widely used in some experiments. Finally, we find that faithful coherent states are distillable and the measurement results provide a lower bound of maximum relative entropy of coherence.

In the rest of this paper, we will introduce some knowledge about coherent states and coherence witness in Sec. \ref{preli}. We investigate the criterion for detecting faithful coherence states under the fidelity-based witness in Sec. \ref{result}, that is, Theorems \ref{Theorem1} and \ref{Theorem2}. In Sec. \ref{applications}, we illustrate the feasibility of unitary transformation experimentally, and study the faithful coherence, coherent distillation and coherence measurement. We conclude in Sec. \ref{conclusion}.

\section{PRELIMINARIES}
\label{preli}
Consider the $d$-dimensional Hilbert space $\mathcal{H}$, for a particular basis $\{|i\rangle\}_{i=1}^{d}$, the incoherent states \cite{TBMC} are defined as those with diagonal density matrices. That is,
\begin{eqnarray}\label{inc1}
\delta=\sum_{i=1}^{d}\delta_{i}|i\rangle\langle i|,
\end{eqnarray}
where $0\leq \delta_{i}\leq 1$ and $\sum_{i=1}^{d}\delta_{i}=1$. We denote $\textbf{I}$ as the set of all incoherent states,
any states $\rho\notin \textbf{I}$ which have non-zero elements $\rho_{ij}$ in non-diagonal blocks are called coherent. Clearly, the set $\textbf{I}$ is convex and compact. Thus, there must exist a hyperplane, namely, coherence witness which separates an arbitrary given coherent state from the set $\textbf{I}$ by the Hahn-Banach theorem \cite{REEF}. A coherence witness $W$ is a Hermitian operator with the property that $\tr(W\delta)\geq 0$ for all incoherence states $\delta\in \textbf{I}$. For any coherent states $\rho$, there exists a witness $W$ such that $\tr(W\rho)<0$. The condition $\tr(W\delta)\geq 0$ can be restricted to $\tr(W\delta)= 0$ for for all incoherence states, and there exists a witness $W$ such that $\tr(W\rho)\neq0$ for any coherent states $\rho$ \cite{HZRA}. However, the former is claimed to be more experimentally operational than the latter \cite{CNTR,MPMC}.

One of the most widespread methods to determine whether a quantum state is coherent, is to measure its fidelity with respect to a pure state. Based on this fact, for a given coherent state $|\Psi\rangle\langle\Psi|$, one has the coherence witness,
\begin{eqnarray}\label{p2}
W=\alpha I-|\Psi\rangle\langle\Psi|,
\end{eqnarray}
where $I$ is the identity matrix and $\alpha=\max\tr(\delta|\Psi\rangle\langle\Psi|)$ is the maximal squared overlap between $|\Psi\rangle$ and the incoherence states $\delta$.  The coherence witness $W$ is a observable which has a positive mean value on all incoherent states. If the observable $W$ ia measured, one obtain $\tr(\varrho W)=\alpha-\langle\Psi|\varrho|\Psi\rangle$, so if $\langle\Psi|\varrho|\Psi\rangle> \alpha$, then the witness detects some coherence. The fidelity-based coherence witness in (\ref{p2}) is easy to construct and can be generalized to the multipartite case.

We say that $\rho$ is unfaithful if it satisfies $\tr(W\rho)\geq 0$ for all witness $W$ of the form (\ref{p2}), i.e., coherence can not be detected by the witness (\ref{p2}). Conversely, a state is faithful if the coherence of it can be detected by the witness (\ref{p2}).

Coherence distillation is a central task in the resource theory of quantum coherence \cite{TBMC,AWDY,KFXW,BRKF}, which  is the process that extracts pure coherence from a mixed state by incoherent operations. For any state, the distillable coherence $C_{d}(\rho)$ is given by the relative entropy of coherence: $C_{d}(\rho)=C_{r}(\rho):=\min_{\sigma\in\Delta}S(\rho\|\sigma)=S(\Delta(\rho))-S(\rho)$, where the decohering operation $\Delta(\rho)=\sum_{i}\langle i|\rho|i\rangle|i\rangle\langle i|$, i. e., the diagonal part of $\rho$ \cite{AWDY}.

Given two operators $\rho$ and $\sigma$ with $\rho\geq0$, $\tr(\rho)\leq1$ and $\sigma\geq0$, the maximum and minimum relative entropy of $\rho$ relative to $\sigma$ are defined as \cite{NDIEEE}
\begin{eqnarray}\label{app1}
D_{\max}(\rho\|\sigma):=\min\{\lambda:\rho\leq2^{\lambda}\sigma\}.
\end{eqnarray}
and
\begin{eqnarray}\label{app2}
D_{\min}(\rho\|\sigma):=-\log\tr(P_{\rho}\sigma),
\end{eqnarray}
where $P_{\rho}$ denotes the projector onto the support of $\rho$. They can define the analogues of the mutual information in the Smooth R$\acute{e}$nyi Entropy framework. The authors of \cite{KBUS} proposed an operational coherence measure based on maximum relative entropy and investigated minimum relative entropy of coherence, respectively, which has some connection with the criterion of faithful coherence we mentioned.

\section{MAIN RESULT}
\label{result}
In this section, we propose the necessary and sufficient conditions for detecting whether the coherent state is faithful in single quantum system and bipartite quantum system, respectively. Namely, Theorem \ref{Theorem1} and Theorem \ref{Theorem2}.

\subsection{Faithful coherence for single systems}
As already mentioned, one possible coherence witness construction makes use of the fact that there are only coherence states around a coherence pure state. So, one can define a coherence witness, $W=\alpha I-|\Psi\rangle\langle\Psi|$,
where $\alpha$ is the maximal squared overlap between $|\Psi\rangle$ and the incoherence states. This can be computed as $\alpha=\max|\langle\Psi|\delta|\Psi\rangle|=\max\sum_{i}\delta_{i}\beta^{2}_{i}=\beta^{2}_{1}$, where the $\beta_{i}$ for $i=1,\ldots, d$ are decreasing ordered nonzero coefficients of the $|\Psi\rangle=\sum_{i=1}^{d}\beta_{i}|i\rangle$ and $\delta_{i}$ for $i=1,\ldots, d$ are shown in (\ref{inc1}). The smallest possible $\alpha$ occurs if the state $|\Psi\rangle$ is a maximally coherence state, for instance $|\Psi\rangle=|\Phi^{+}\rangle=\sum_{i=1}^{d}\frac{1}{\sqrt{d}}|i\rangle$. Then we have
\begin{eqnarray}\label{w2}
\mathcal{W}=\frac{I}{d}-|\Phi^{+}\rangle\langle\Phi^{+}|.
\end{eqnarray}
Clearly, there are many other maximally coherence states which are equivalent to $|\Phi^{+}\rangle$. For reasons that become apparent later, we call this set of witness of the type in Eq. (\ref{w2}) the relevant fidelity coherence witnesses (RFCW).

\bt
\label{Theorem1}
Let $\rho$ be a faithful coherent state, i.e., its coherence can be detected by some fidelity-based coherence witness in (\ref{p2}). Then $\rho$ can be detected by a RFCW. In other words, a state $\rho$ is faithful if and only if there is a unitary transformation $U$ such that
 \begin{eqnarray}\label{Th1}
\langle\Phi^{+}|U^{\dagger}\rho U|\Phi^{+}\rangle > \frac{1}{d}.
\end{eqnarray}
\et

\bpf
Assume that $\rho$ can be detected dy the fidelity-based coherence witness $W=\alpha I-|\Psi\rangle\langle\Psi|$, with $|\Psi\rangle=\sum_{i=1}^{d}\beta_{i}|i\rangle$. We consider $2^{d-1}$ RFCWs, coming from the maximally coherent states
\begin{eqnarray}\label{th11}
|\phi_{\vec{a}}\rangle=\frac{1}{\sqrt{d}}(|1\rangle+\sum_{j=2}^{d}a_{j}|j\rangle),
\end{eqnarray}
where $\vec{a}=(a_{2},a_{3},\ldots, a_{d})$ and the coefficients $a_{j}$ can have possible values $\pm1$. This leads to $2^{d-1}$ RFCWs $\mathcal{W}_{\vec{a}}=\frac{I}{d}-|\phi_{\vec{a}}\rangle\langle\phi_{\vec{a}}|$.

Our aim is to find probabilities $p_{\vec{a}}$ such that the operator
\begin{eqnarray}\label{th12}
\mathcal{D}=W-d\beta_{1}^{2}\sum_{\vec{a}}p_{\vec{a}}\mathcal{W}_{\vec{a}}
\end{eqnarray}
is positive semidefinite. Thus, $\tr(\rho W)< 0$ implies that there at least exists one $\vec{a}$ such that one RFCW $\mathcal{W}_{\vec{a}}$ satisfies $\tr(\rho \mathcal{W}_{\vec{a}})< 0$. By direct calculation, we find that the diagonal entries of operator $\mathcal{D}$ is independent of $p_{\vec{a}}$, and the value of diagonal entries is either $0$ or $\beta_{1}^{2}-\beta_{i}^{2}\geq 0$. In this case, if $\mathcal{D}$ is a diagonal operator, it can be ensured that $\mathcal{D}\geq 0$. To make $\mathcal{D}$ diagonal, one need to find probabilities $p_{\vec{a}}$ such that the off-diagonal entries satisfy
\begin{eqnarray}\label{th13}
&&\sum_{\vec{a}}p_{\vec{a}}\left(
                           \begin{array}{ccccc}
                             \ast & a_{2} & a_{3}& \ldots & a_{d} \\
                             a_{2} & \ast & a_{2}a_{3}& \ldots & a_{2}a_{d} \\
                             a_{3} & a_{2}a_{3} & \ast & \ldots & a_{3}a_{d} \\
                             \vdots & \vdots & \vdots& \ddots & \vdots \\
                             a_{d} & a_{2}a_{d} & a_{3}a_{d}& \ldots & \ast \\
                           \end{array}
                         \right)\nonumber\\
&&=\left(
     \begin{array}{ccccc}
       \ast &\gamma_{2} & \gamma_{3}& \ldots & \gamma_{d} \\
       \gamma_{2} & \ast & \gamma_{2}\gamma_{3}& \ldots & \gamma_{2}\gamma_{d} \\
       \gamma_{3} & \gamma_{2}\gamma_{3} & \ast & \ldots & \gamma_{3}\gamma_{d} \\
       \vdots & \vdots & \vdots& \ddots & \vdots \\
       \gamma_{d} & \gamma_{2}\gamma_{d} & \gamma_{3}\gamma_{d}& \ldots & \ast \\
     \end{array}
   \right),
\end{eqnarray}
where $\gamma_{i}=\frac{\beta_{i}}{\beta_{1}}\in [0, 1]$ for $i=2, 3, \ldots, d$. We can view the $\gamma_{i}$ as expectation values of some observables (such as $\sigma_{x}$) on a $(d-1)$ qubits product state $\varrho=\varrho_{1}\otimes \varrho_{2}\otimes \ldots \otimes \varrho_{d-1}$. The term $\gamma_{i}\gamma_{j}$ correspond to two-body correlation $\langle\sigma_{x}^{i-1}\otimes\sigma_{x}^{j-1}\rangle$ on the same state \cite{OGYM}. On the other hand, the left hand of (\ref{th13}) can be seen as a local hidden variable model \cite{RFWQ}, where the index $\vec{a}$ is the hidden variable occurring with probability $p_{\vec{a}}$, and the $a_{i}$ are the deterministic assignments for the measurement results of $\sigma_{x}$ on the different particles. For fully separable states, it is well known that all measurements can be explained by a local hidden variable model. \cite{RFWQ} Therefore, there must be $p_{\vec{a}}$ so that equation (\ref{th13}) holds.

In other words, if a state $\rho$ is faithful, then it can be detected by a coherence witness in RFCW. That is, there exists a unitary transformation $U$ such that
\begin{eqnarray}\label{th15}
\tr(\rho\mathcal{W}_{U})<0,
\end{eqnarray}
with $\mathcal{W}_{U}=\frac{I}{d}-U^{\dagger}|\Phi^{+}\rangle\langle\Phi^{+}|U$. The above inequality can be calculated
$\langle\Phi^{+}|U\rho U^{\dagger}|\Phi^{+}\rangle >  \frac{1}{d}$. Conversely, if a quantum coherent state can be detected by RFCW, it must be faithful.
$\square$
\epf

From Theorem \ref{Theorem1}, it is immediately clear that already for single-qubit not all coherent states are faithful for a fixed basis. We know that a single qubit state can be written as
\begin{eqnarray}\label{sq1}
\rho=\frac{1}{2}(I+\sum_{i=1}^{3}s_{i}\sigma_{i}),
\end{eqnarray}
with Pauli matrices $\sigma_{i}$, and the Bloch vector $\vec{s}=(s_{1}, s_{2}, s_{3})$. According to Theorem \ref{Theorem1}, a single qubit state $\rho$ is faithful if and only if $0< s_{1}< 1$. Thus, the states with $s_{1}=0$ or $s_{1}=1$ are clearly not faithful. Especially for $s_{1}=1$, the single qubit state is reduced to state
$\frac{1}{2}\left(
              \begin{array}{cc}
                1 & 0 \\
                0 & 1 \\
              \end{array}
            \right)
$, which is an incoherent state.

\begin{widetext}
In Theorem \ref{Theorem1}, the unitary transformations $U$ form a set $\textbf{U}$,
\begin{eqnarray}\label{Uset}
\textbf{U}=\{U|&&U\left(
                 \begin{array}{c}
                   1 \\
                   1 \\
                   1 \\
                   \vdots \\
                   1 \\
                 \end{array}
               \right)=\left(
                         \begin{array}{c}
                           1 \\
                           -1 \\
                           1 \\
                           \vdots \\
                           1 \\
                         \end{array}
                       \right), \quad \rm or \ldots, \quad \rm or \quad U\left(
                 \begin{array}{c}
                   1 \\
                   1 \\
                   1 \\
                   \vdots \\
                   1 \\
                 \end{array}
               \right)=\left(
                         \begin{array}{c}
                           1 \\
                           1 \\
                           1 \\
                           \vdots \\
                           -1 \\
                         \end{array}
                       \right), \quad \rm or \nonumber\\
                       &&U\left(
                 \begin{array}{c}
                   1 \\
                   1 \\
                   1 \\
                   \vdots \\
                   1 \\
                 \end{array}
               \right)=\left(
                         \begin{array}{c}
                           1 \\
                           -1 \\
                           -1 \\
                           \vdots \\
                           1 \\
                         \end{array}
                       \right), \quad \rm or \ldots, \quad \rm or \quad U\left(
                 \begin{array}{c}
                   1 \\
                   1 \\
                   1 \\
                   \vdots \\
                   1 \\
                 \end{array}
               \right)=\left(
                         \begin{array}{c}
                           1 \\
                           -1 \\
                           1 \\
                           \vdots \\
                           -1 \\
                         \end{array}
                       \right), \quad \rm or \nonumber\\
                       && \vdots \nonumber\\
                       &&U\left(
                 \begin{array}{c}
                   1 \\
                   1 \\
                   1 \\
                   \vdots \\
                   1 \\
                 \end{array}
               \right)=\left(
                         \begin{array}{c}
                           1 \\
                           -1 \\
                           -1 \\
                           \vdots \\
                           -1 \\
                         \end{array}
                       \right)
\},
\end{eqnarray}
which contains $\frac{d(d-1)}{2}$ equations and makes the experimental realization of Theorem \ref{Theorem1} operational. Because $U$ satisfying each equation is not unique, the number of elements in set $\textbf{U}$ is greater than $\frac{d(d-1)}{2}$. Nevertheless, we only need to choose a $U$ in each equation to realize the inequality (\ref{Th1}) in Theorem \ref{Theorem1}. The specific discussion will be presented in section \ref{experimental}.
\end{widetext}
\subsection{Faithful coherence for bipartite systems}
Quantum coherence can be generalized to the multipartite system \cite{TRBM}. Take bipartite system as an example, the bipartite incoherence states is the convex combinations of states of the form $|k\rangle|l\rangle$, with $\{|k\rangle\}$ and $\{|l\rangle\}$ the incoherent reference bases for each subsystem, respectively. That is,
\begin{eqnarray}\label{inc2}
\hat{\delta}=\sum_{k,l=1}^{d}\hat{\delta}_{kl}|kl\rangle\langle kl|,
\end{eqnarray}
with $0\leq \hat{\delta}_{kl}\leq 1$ , $\sum_{k,l=1}^{d}\hat{\delta}_{kl}=1$. One commonly fidelity-based coherence witness has the following form,
\begin{eqnarray}\label{W1}
\hat{W}=\hat{\alpha} I-|\hat{\Psi}\rangle\langle\hat{\Psi}|,
\end{eqnarray}
where $I$ is the identity matrix,  $|\hat{\Psi}\rangle$ is a given coherent state and $\hat{\alpha}$ is the maximal squared overlap between $|\hat{\Psi}\rangle$ and the incoherence states $\hat{\delta}$. The smallest possible $\hat{\alpha}$ occurs if the state $|\hat{\Psi}\rangle$ is a maximally coherence state, for instance $|\hat{\Psi}\rangle=|\hat{\Phi^{+}}\rangle=\frac{1}{\sqrt{d}}\sum_{i,j=1}^{d}|ij\rangle$. Then we have
\begin{eqnarray}\label{W2}
\hat{\mathcal{W}}=\frac{I}{d}-|\hat{\Phi^{+}}\rangle\langle\hat{\Phi^{+}}|.
\end{eqnarray}
There are many other bipartite maximally coherence states which are locally equivalent to $|\hat{\Phi^{+}}\rangle$. For reasons, we call the set of witnesses in Eq. (\ref{W2}) the bipartite relevant fidelity coherence witnesses (BRFCW). Then we obtain a result similar to Theorem \ref{Theorem1}.

\bt
\label{Theorem2}
Let $\rho_{AB}$ be a faithful coherent state, i.e., its coherence can be detected by some fidelity-based coherence witness in (\ref{W1}). Then $\rho_{AB}$ can be detected by a relevant fidelity coherence witness in (\ref{W2}). In other words, a state $\rho_{AB}$ is faithful if and only if there are local unitary transformations $U_{A}$ and $U_{B}$ such that
 \begin{eqnarray}\label{Th2}
\langle\hat{\Phi^{+}}|U_{A}^{\dagger}\otimes U_{B}^{\dagger}\rho_{AB} U_{A}\otimes U_{B}|\hat{\Phi^{+}}\rangle > \frac{1}{d}.
\end{eqnarray}
\et

\begin{figure}[h]
\centering
\includegraphics[width=3.5in]{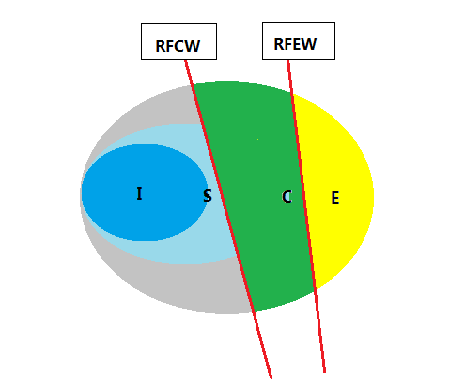}
\caption{For bipartite quantum states, faithful entangled states must be faithful coherent states. Here, RFEW means relevant fidelity entanglement witnesses in \cite{OGYM}. On the left of RECW, the green and yellow parts represent the faithful coherent states, and on the left of RFEW, the yellow part represents the faithful entangled states.}
\label{fig:RFECW}
\end{figure}

The proof of Theorem \ref{Theorem2} follows from the proof in Theorem \ref{Theorem1}. Ref \cite{OGYM} showed that not all bipartite entangled states are faithful. Thus, bipartite faithful entangled states must be faithfully coherent states, which is shown in FIG \ref{fig:RFECW}.

\section{APPLICATIONS}
\label{applications}

In this section, we investigate two parts: one is the experimental realizations of Theorems \ref{Theorem1} and \ref{Theorem2}, and the other is the relation between faithful coherence and coherence distillation.

\subsection{Experimental explanation of unitary transformation}
\label{experimental}
\begin{figure}[h]
\centering
\includegraphics[width=3.0in]{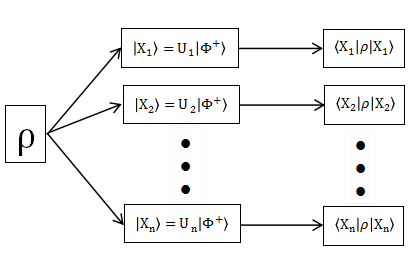}
\caption{For a quantum state, a set of values can be obtained by inner product with a finite number of $X_{i}$. Here $n=\frac{d(d-1)}{2}$.}
\label{fig:uset}
\end{figure}

In order to realize the unitary transformation in Theorem \ref{Theorem1} experimentally, we can find a suitable solution for each equation in set $\textbf{U}$ of (\ref{Uset}), so that it contains only $\frac{d(d-1)}{2}$ elements. To detect whether a quantum state is faithful, the experimental process can be seen in FIG \ref{fig:uset}. In the set of values obtained, if there is a value greater than $\frac{1}{d}$, then the $\rho$ is faithful, and the unitary transformation $U$ corresponding to this value can be realized experimentally. If all values are less than $\frac{1}{d}$, then $\rho$ is unfaithful.

\begin{figure}[h]
\centering
\includegraphics[width=2.0in]{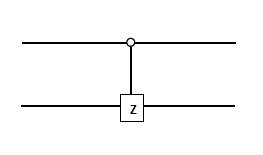}
\caption{ When the first qubit is set to zero, controlled operation with a Pauli-Z gate is performed on the second qubit.}
\label{fig:TQU11}
\end{figure}

Next, we will discuss the experimental feasibility of satisfying the unitary transformation $U$ in set (\ref{Uset}). These $\frac{d(d-1)}{2}$ elements can form a matrix in following form, which can be regarded as the matrix form of set $\textbf{U}$,
\begin{eqnarray}\label{upper}
\textbf{U}_{matrix}=\left(
                      \begin{array}{ccccc}
                        U_{11} & U_{12} & \cdots & U_{1,d-2} & U_{1,d-1} \\
                        U_{21} & U_{22} & \cdots & U_{2,d-2} &   \\
                        U_{31} & U_{32} & \cdots &   &   \\
                        \vdots &   &   &   &   \\
                        U_{d-1,1} &   &   &   &   \\
                      \end{array}
                    \right),\nonumber\\
\end{eqnarray}
where $U_{ij}$ is a solution of the equation corresponding to the corresponding position in the set (\ref{Uset}). For example, $U_{11}$ satisfying the first equation in set (\ref{Uset}) can be selected as
\begin{eqnarray*}
U_{11}=\left(
        \begin{array}{ccccc}
          1 & 0 & 0 & \cdots & 0 \\
          0 & -1 & 0 & \cdots & 0 \\
          0 & 0 & 1 & \cdots & 0 \\
          \vdots & \vdots & \vdots & \vdots & \vdots \\
          0 & 0 & 0 & \cdots & 1 \\
        \end{array}
      \right),
\end{eqnarray*}
$U_{12}$ satisfying the second equation in set (\ref{Uset}) can be selected as
\begin{eqnarray*}
U_{12}=\left(
        \begin{array}{ccccc}
          1 & 0 & 0 & \cdots & 0 \\
          0 & 1 & 0 & \cdots & 0 \\
          0 & 0 & -1 & \cdots & 0 \\
          \vdots & \vdots & \vdots & \vdots & \vdots \\
          0 & 0 & 0 & \cdots & 1 \\
        \end{array}
      \right),
\end{eqnarray*}
and
\begin{eqnarray*}
U_{1,d-1}=\left(
        \begin{array}{ccccc}
          1 & 0 & 0 & \cdots & 0 \\
          0 & 1 & 0 & \cdots & 0 \\
          0 & 0 & 1 & \cdots & 0 \\
          \vdots & \vdots & \vdots & \vdots & \vdots \\
          0 & 0 & 0 & \cdots & -1 \\
        \end{array}
      \right),
\end{eqnarray*}
\begin{eqnarray*}
U_{21}=\left(
        \begin{array}{ccccc}
          1 & 0 & 0 & \cdots & 0 \\
          0 & -1 & 0 & \cdots & 0 \\
          0 & 0 & -1 & \cdots & 0 \\
          \vdots & \vdots & \vdots & \vdots & \vdots \\
          0 & 0 & 0 & \cdots & 1 \\
        \end{array}
      \right),
\end{eqnarray*}
\begin{eqnarray*}
U_{2,d-1}=\left(
        \begin{array}{ccccc}
          1 & 0 & 0 & \cdots & 0 \\
          0 & -1 & 0 & \cdots & 0 \\
          0 & 0 & 1 & \cdots & 0 \\
          \vdots & \vdots & \vdots & \vdots & \vdots \\
          0 & 0 & 0 & \cdots & -1 \\
        \end{array}
      \right),
\end{eqnarray*}
\begin{eqnarray*}
U_{d-1,1}=\left(
        \begin{array}{ccccc}
          1 & 0 & 0 & \cdots & 0 \\
          0 & -1 & 0 & \cdots & 0 \\
          0 & 0 & -1 & \cdots & 0 \\
          \vdots & \vdots & \vdots & \vdots & \vdots \\
          0 & 0 & 0 & \cdots & -1 \\
        \end{array}
      \right),
\end{eqnarray*}
In this way, we get the $\frac{d(d-1)}{2}$ elements in the set $\textbf{U}$ completely. The element in row $i (2\leq i\leq d-1)$ of $\textbf{U}_{matrix}$ is the product of some $(i-1)$ elements in the first row and $U_{11}$. We know that any unitary matrix specifies a valid quantum gate. A quantum computer is built from a quantum circuit containing wires and elementary quantum gates to carry around and manipulate the quantum information. Thus, we can realize the unitary transformation in Theorem \ref{Theorem1} by using quantum gates and circuits. Here we might as well make the dimension of $U_{ij}$ $d=2^{k} (k=1, 2, 3, \ldots)$, so we can use multiple qubit gates to realize unitary transformation. Firstly, we study the circuit process of realizing $U_{11}$ in the case of two qubits (i.e. $k = 2$),
\begin{eqnarray*}
U_{11}|_{k=2}&&=\left(
         \begin{array}{cccc}
           1 & 0 & 0 & 0 \\
           0 & -1 & 0 & 0 \\
           0 & 0 & 1 & 0 \\
           0 & 0 & 0 & 1 \\
         \end{array}
       \right)\nonumber\\
       &&=|0\rangle\langle0|\otimes \sigma_{z}+|1\rangle\langle1|\otimes I_{2},
\end{eqnarray*}
where $\sigma_{z}$ is Pauli-Z gate and $I_{2}$ is $2\ast 2$ identity matrix. Quantum circuit are shown in FIG \ref{fig:TQU11}.

\begin{figure}[h]
\centering
\includegraphics[width=2.0in]{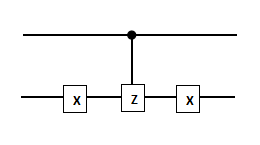}
\caption{ Controlled operation when the first qubit is set to one.}
\label{fig:TQU12}
\end{figure}

Similarly, $U_{12}|_{k=2}$ can be realized by circuit in FIG \ref{fig:TQU12}.
\begin{eqnarray*}
U_{12}|_{k=2}=\left(
         \begin{array}{cccc}
           1 & 0 & 0 & 0 \\
           0 & 1 & 0 & 0 \\
           0 & 0 & -1 & 0 \\
           0 & 0 & 0 & 1 \\
         \end{array}
       \right).
\end{eqnarray*}

\begin{figure}[h]
\centering
\includegraphics[width=2.0in]{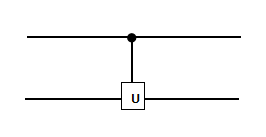}
\caption{ Controlled-U  operation}
\label{fig:TQU31}
\end{figure}

The last $U_{31}|_{k=2}$ can be realized by circuit in FIG \ref{fig:TQU31}.
\begin{eqnarray*}
U_{31}|_{k=2}=\left(
         \begin{array}{cccc}
           1 & 0 & 0 & 0 \\
           0 & -1 & 0 & 0 \\
           0 & 0 & -1 & 0 \\
           0 & 0 & 0 & -1 \\
         \end{array}
       \right).
\end{eqnarray*}

\begin{figure}[h]
\centering
\includegraphics[width=2.0in]{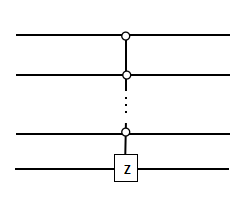}
\caption{ When the first $k-1$ qubits are set to zero, controlled operation with a Pauli-Z gate is performed on the last qubit.}
\label{fig:TQUk11}
\end{figure}

\begin{figure}[h]
\centering
\includegraphics[width=2.0in]{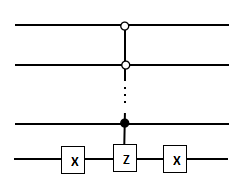}
\caption{Controlled operation when the first $k-2$ qubits are set to zero and the $(k-1)$-th qubit is set to one.}
\label{fig:TQUk12}
\end{figure}

\begin{figure}[h]
\centering
\includegraphics[width=2.0in]{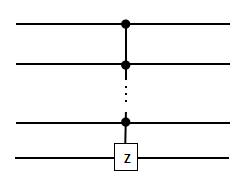}
\caption{Controlled operation when the first $k-1$ qubits are set to one.}
\label{fig:TQUk12k}
\end{figure}

Moreover, we can also realize the unitary transformation under multiple qubits ($k>2$). $U_{11}$ can be written as
\begin{eqnarray*}
U_{11}|_{k}&&=|\underbrace{00\ldots0}_{k-1}\rangle\langle\underbrace{00\ldots0}_{k-1}|\otimes \sigma_{z}\nonumber\\
&&+|00\ldots1\rangle\langle00\ldots1|\otimes I_{2}\nonumber\\
&&+\ldots+|11\ldots1\rangle\langle11\ldots1|\otimes I_{2},
\end{eqnarray*}
and Quantum circuit are shown in FIG \ref{fig:TQUk11}.
$U_{12}|_{k}$ can be written as
\begin{eqnarray*}
U_{12}|_{k}&&=|\underbrace{00\ldots0}_{k-1}\rangle\langle\underbrace{00\ldots0}_{k-1}|\otimes I_{2}\nonumber\\
&&+|00\ldots1\rangle\langle00\ldots1|\otimes (-\sigma_{z})\nonumber\\
&&+\ldots+|11\ldots1\rangle\langle11\ldots1|\otimes I_{2},
\end{eqnarray*}
and Quantum circuit are shown in FIG \ref{fig:TQUk12}.
$U_{1,2^{k}}|_{k}$ can be written as
\begin{eqnarray*}
U_{1,2^{k}}|_{k}&&=|\underbrace{00\ldots0}_{k-1}\rangle\langle\underbrace{00\ldots0}_{k-1}|\otimes I_{2}\nonumber\\
&&+|00\ldots1\rangle\langle00\ldots1|\otimes I_{2}\nonumber\\
&&+\ldots+|11\ldots1\rangle\langle11\ldots1|\otimes \sigma_{z},
\end{eqnarray*}
and Quantum circuit are shown in FIG \ref{fig:TQUk12k}.

According to the above process, we can get more quantum circuits, so that we can realize the unitary transformation experimentally. In this way, Theorem \ref{Theorem1} can be operational and make future experiments easier to implement. It is worth mentioning that the unitary transformation in Theorem \ref{Theorem2} is the tensor product of the unitary transformation on the two subsystems, which can be studied by similar methods.

\subsection{Coherence distillation and coherence measurement}
\label{distillation}
In \cite{AWDY}, there is no bound coherence ,that is , every state with any coherence is distillable. Thus, faithful coherence states are distillable.

The maximum relative entropy of coherence is defined as \cite{KBUS}
\begin{eqnarray}\label{app3}
C_{\max}(\rho):=\min_{\sigma\in \textbf{I}}D_{\max}(\rho\|\sigma).
\end{eqnarray}
And they showed that $2^{C_{\max}}$ is equal to the maximum overlap with the maximally coherent state that can be achieved by different types of free operations DIO, IO and SIO, i. e.,
\begin{eqnarray}\label{app4}
2^{C_{\max}(\rho)}&&=d\max_{\varepsilon,|\Phi^{+}\rangle}F[\varepsilon(\rho),|\Phi^{+}\rangle\langle\Phi^{+}|]^{2}\nonumber\\
&&=d\max_{\varepsilon,|\Phi^{+}\rangle}\tr[\rho\varepsilon^{\dagger}(|\Phi^{+}\rangle\langle\Phi^{+}|)],
\end{eqnarray}
where $F(\rho,\sigma)=\tr[|\sqrt{\rho}\sqrt{\sigma}|]$is the fidelity between states $\rho$ and $\sigma$, $|\Phi^{+}\rangle$ is the maximally coherent state, and $\varepsilon$ belongs to either DIO, IO or SIO.

We apply the RFCW to any quantum state $\rho$, i. e., $\tr(\mathcal{W}\rho)=\frac{1}{d}-\tr[\rho |\Psi\rangle\langle\Psi|]$, then we get
\begin{eqnarray}\label{app5}
1-d\tr(\mathcal{W}\rho)=d \tr[\rho|\Phi^{+}\rangle\langle\Phi^{+}|].
\end{eqnarray}

When $\varepsilon$ belongs to DIO, we know the identity operation $\textbf{I}(\cdot)=(\cdot)$  also belongs to DIO, thus
\begin{eqnarray}\label{app6}
d \tr[\rho|\Phi^{+}\rangle\langle\Phi^{+}|]\leq d\max_{\varepsilon,|\Phi^{+}\rangle}\tr[\rho\varepsilon^{\dagger}(|\Phi^{+}\rangle\langle\Phi^{+}|)],\nonumber\\
\end{eqnarray}
equivalently,
\begin{eqnarray}\label{app7}
1-d\tr(\mathcal{W}\rho)\leq 2^{C_{\max}(\rho)}.
\end{eqnarray}
Similar results can be obtained when $\varepsilon$ belongs to either IO or SIO. We find that the measurement result of RFCW for quantum states gives a lower bound of maximum relative entropy of coherence $C_{\max}(\rho)$. The measurements do not need a complete tomography of quantum states and is feasible experimentally.

\section{CONCLUSION}
\label{conclusion}
In view of the study of faithful entanglement, we proposed the notion of faithful coherence state based on the fidelity-based coherence witness. We constructed the criterion for detecting the states, which are consistent with a subclass of fidelity-based criterion by using unitary transformations for single and bipartite systems. We have realized these unitary transformations by quantum gates and circuits, and find some relationship of faithful coherence and coherence distillation, and coherence measures. The method used in this paper can also be generalized to arbitrary multipartite qudit systems. It is worth mentioning that the concept of faithful can also be studied in other resource theories, such as quantum thermodynamics \cite{CSLC}.

\bigskip
\noindent{\bf Acknowledgments}\, Authors were supported by the NNSF of China (Grant No. 11871089), and the Fundamental
Research Funds for the Central Universities (Grant Nos. KG12080401 and ZG216S1902).

\end{document}